\documentclass[12pt,a4paper]{article}
\usepackage{natbib,graphicx}

\begin{document}

{\it
\noindent
Melnikov A.V., Shevchenko I.I.\\
``On the rotational dynamics of Prometheus and Pandora''\\
Celestial Mechanics and Dynamical Astronomy.
2008. V.101. N1-2. P.31-47.
}

\vskip2cm

\begin{center}
{\bf \Large On the rotational dynamics of\\ Prometheus and Pandora}\\

\bigskip

{\it A. V. Melnikov\, and I. I. Shevchenko \\
Pulkovo Observatory of the Russian Academy of Sciences,\\
Pulkovskoje ave. 65/1, St.Petersburg 196140, Russia}

\end{center} 
%\maketitle

\begin{abstract}
Possible rotation states of two satellites of Saturn, Prometheus
(S16) and Pandora~(S17), are studied by means of numerical
experiments. The attitude stability of all possible modes of
synchronous rotation and the motion close to these modes is
analyzed by means of computation of the Lyapunov spectra of the
motion. The stability analysis confirms that the rotation of
Prometheus and Pandora might be chaotic, though the possibility of
regular behaviour is not excluded. For the both satellites, the
attitude instability zones form series of concentric belts
enclosing the main synchronous resonance center in the phase space
sections. A hypothesis is put forward that these belts might form
``barriers'' for capturing the satellites in synchronous rotation.
The satellites in chaotic rotation can mimic ordinary regular
synchronous behaviour: they preserve preferred orientation for
long periods of time, the largest axis of satellite's figure being
directed approximately towards Saturn.
\end{abstract}

\section{Introduction}
\label{intro}

In 1980s, \cite{WPM84} and \cite{W87} demonstrated theoretically
that a planetary satellite of non-spherical shape in an elliptic
orbit can rotate in a chaotic, unpredictable way. They found that
the most probable candidate for the chaotic rotation, due to
pronounced shape asymmetry and significant orbital eccentricity,
was the satellite of Saturn Hyperion~(S7). Later on, a direct
modelling of observed light curves of Hyperion~\citep{K89a, K89b,
TBN95, BNT95, DGG02} confirmed the chaotic character of its
rotation.

Recently it was found in a theoretical research~\citep{KS05}, that
two other satellites of Saturn, Prometheus~(S16) and
Pandora~(S17), can also reside in a state of chaotic rotation.
Contrary to the case of Hyperion, chaos in rotation of these two
satellites is due to fine-tuning of the dynamical and physical
parameters rather than simply to a large extent of a chaotic zone
in the rotational phase space.

It is remarkable that the orbital dynamics of Prometheus and
Pandora are known to be chaotic with the Lyapunov time of only
three years, and this dynamical chaos is directly observed;
see~\citep{GR03a,GR03b,CM04,FG06}. However note that the
theoretical inferences on chaos in rotation of these satellites
are completely independent from the existence of orbital chaos.

In the present paper we study the problem of rotational dynamics
of these two satellites in detail, exploring the attitude
stability not only in the centers of synchronous resonances, but
also in the phase space in the vicinities of the resonances. Our
analysis includes period-doubling bifurcation modes of synchronous
spin-orbit states. We take into account all available modern
observational data.

An important problem, first of all from the observational point of
view, is the following: whether there exists a preferred
orientation of the satellites rotating chaotically, or all
orientations are equiprobable? The orientation of a satellite in
chaotic rotation, generally speaking, is not necessarily
isotropic. This was demonstrated in a numerical experiment
by~\cite{W87} in calculating the rotation of Phobos in the
vicinities of the separatrices of the 1:2 spin-orbit resonance;
see Fig.~5 in \citep{W87}.

The plan of the present work is as follows. First, we study the
attitude stability of the planar rotation of Prometheus and
Pandora with respect to tilting the axis of rotation. The
stability analysis is carried out for all possible exact modes of
synchronous resonance ($\alpha$-resonance, $\beta$-resonance, and
the period-doubling bifurcation mode of $\alpha$-resonance), as
well as for the trajectories of all possible kinds (periodic,
quasiperiodic, chaotic in the planar problem) on a representative
section of the phase space of planar rotation. Then the problem on
the preferred orientation of Prometheus and Pandora in chaotic
rotation is considered.

\section{The reference frame and the equations of motion}
\label{sec:syseq}

We suppose that a satellite represents a non-spherical rigid body
moving around a planet (a gravitating point) in a non-perturbed
elliptic orbit with the eccentricity $e$. Location of the
satellite in the orbit is determined by the true anomaly $f$ or
the eccentric anomaly $E$. The shape of the satellite is described
by a triaxial ellipsoid with the principal semiaxes $a > b > c$
and the corresponding principal central moments of inertia $A < B
< C$. The size of the satellite is much less than the radius of
the orbital motion. The dynamics of the three-dimensional rotation
of the satellite is determined by the parameters $e$, $b/a$, $c/b$
and the initial conditions of the motion. The angular velocities
of rotation are expressed in the units of the orbital mean motion,
and the ``satellite~--- planet'' distance is expressed in the
units of the semimajor axis of the orbit. One orbital period
corresponds to $2\pi$ time units.

The oblateness of Saturn causes precession of the pericenters of
the orbits of the satellites, the precession rate being equal to
$3.1911 \times 10^{-5}$ $^\circ s^{-1}$ in the case of Prometheus
and $3.0082 \times 10^{-5}$ $^\circ s^{-1}$ in the case of Pandora
\citep{GR03b}. Hence the periods of precession are equal to 0.36
and 0.38 yr, respectively. This is much greater that the Lyapunov
times (less than 1 d; see \cite{KS05}) of the rotation of these
satellites, if it were chaotic. Since the timescales are so
different, we ignore the precession of orbits in our study.

For describing the orientation of the satellite we use an inertial
frame identical to that used in~\citep{WPM84, MS98, MS00}. This
$Oxyz$ frame is defined initially at the pericenter of the orbit
as follows: the $x$ axis is directed along the ``orbit
pericenter~--- planet'' vector, the $y$ axis is parallel to the
vector of the orbital velocity at the pericenter, the $z$ axis is
orthogonal to the orbital plane and completes the reference system
to a right-handed system. Orientation of the satellite with
respect to the axes of the $Oxyz$ frame is defined by a sequence
of imaginary rotations of the satellite by the Euler angles
$\theta$, $\phi$, $\psi$ from an initial position until the
satellite reaches its actual orientation. In the initial position
the axes $a$, $b$, $c$ coincide with the axes $x$, $y$, $z$,
respectively. The axes $a$, $b$, $c$ are directed along the
principal axes of inertia with the moments $A$, $B$, $C$,
respectively, and are ``frozen'' in the satellite. The imaginary
rotations are carried out in the following sequence: first,
rotation by $\theta$ about $c$, second, rotation by $\phi$ about
$a$, and third, rotation by $\psi$ about $b$.

The definition of the Euler angles adopted here is identical to
that described and used by \citet{WPM84} and is different from the
usual one. The reason for using the alternative Euler angle set is
that the standard one has a coordinate singularity at the point
where a satellite's $c$ axis (the axis of the maximum moment of
inertia) is orthogonal to the orbit plane. In the adopted frame
this position corresponds to $\phi = 0$, while the singularity is
shifted to $\phi = \pm{\pi/2}$. The latter value corresponds to
the satellite's axis of rotation lying in the orbit plane.

Rotation of the satellite is described by Euler's dynamic and
kinematic equations. The dynamic equations~\citep{B65, WPM84} can
be written as

\begin{equation}
\label{eyleq}
\left\{
\begin{array}{lcl}
{\displaystyle
A{\mathrm{d}\omega_a \over \mathrm{d}t}-\omega_b\omega_c(B-C)}
& = &
{\displaystyle
{}-3{{GM}\over{r^3}}\beta\gamma(B-C),}
\cr\cr
{\displaystyle
B{\mathrm{d}\omega_b \over \mathrm{d}t}-\omega_c\omega_a(C-A)}
& = &
{\displaystyle
{}-3{{GM}\over{r^3}}\gamma\alpha(C-A),}
\cr\cr
{\displaystyle
C{\mathrm{d}\omega_c \over \mathrm{d}t}-\omega_a\omega_b(A-B)}
& = &
{\displaystyle
{}-3{{GM}\over{r^3}}\alpha\beta(A-B).}
\end{array}
\right.
\end{equation}

\noindent Here $G$ is the gravitational constant; $M$ is the mass
of the planet; $\omega_a$, $\omega_b$, $\omega_c$ are the
projections of the vector of the angular velocity ${\vec\omega}$
on the axes $a$, $b$, $c$; $r = a (1 - e \cos E)$ is the
``satellite~--- planet'' distance, $a$ is the semimajor axis of
the orbit; $\alpha$, $\beta$, $\gamma$ are the direction cosines
of the principal axes of inertia with respect to the direction to
the planet.

The kinematic equations and expressions for direction cosines, in
the reference frame used here, according to~\citep{WPM84, MS98},
are given by

\begin{equation}
\label{angvel_1}
\left\{
\begin{array}{lcl}
\omega_a & = &
{\displaystyle
{\mathrm{d}\theta \over \mathrm{d}t}\sin\phi\sin\psi+
           {\mathrm{d}\phi \over \mathrm{d}t}\cos\psi,
}
\cr\cr
\omega_b & = &
{\displaystyle
{\mathrm{d}\theta \over \mathrm{d}t}\sin\phi\cos\psi-
           {\mathrm{d}\phi \over \mathrm{d}t}\sin\psi,
}
\cr\cr
\omega_c & = &
{\displaystyle
{\mathrm{d}\theta \over \mathrm{d}t}\cos\phi+
{\mathrm{d}\psi \over \mathrm{d}t},}
\end{array}
\right.
\end{equation}

\begin{equation}
\label{dircos_1}
\left\{
\begin{array}{lcl}
\alpha & = & \cos(\theta-f)\cos\psi-\sin(\theta-f)\cos\phi\sin\psi,\cr
\beta  & = & -\cos(\theta-f)\sin\psi-\sin(\theta-f)\cos\phi\cos\psi,\cr
\gamma & = & \sin(\theta-f)\cos\phi.
\end{array}
\right.
\end{equation}

\noindent Eqs.~(\ref{eyleq}), (\ref{angvel_1}), and
(\ref{dircos_1}) are used in what follows for calculation of the
rotational dynamics.

\section{Synchronous resonance regimes and attitude stability of rotation}
\label{sec:stab}

\subsection{The ``Amalthea effect''}
\label{sec:amal}

In~\citep{MS98, MS00} it was found for Amalthea~(J5) that two
different synchronous regimes of rotation, ``$\alpha$-resonance''
and ``$\beta$-resonance'', coexist in the phase space of rotation
of this satellite. By means of calculation of the Lyapunov
characteristic exponents (LCEs), \cite{MS98} showed that the
planar rotation of Amalthea in the $\beta$-resonance center and
its neighbourhood was stable with respect to tilting the axis of
rotation, while its rotation in the $\alpha$-resonance center and
its neighbourhood was unstable. (Note that the terms
``$\alpha$-resonance'' and ``$\beta$-resonance'' had not yet been
used in that paper.) In~\citep{MS00} by means of calculation and
statistical analysis of the multipliers of the periodic solutions
corresponding to the $\alpha$- and $\beta$-resonances, the
stability of rotation of Amalthea at the centers of the both modes
was investigated on a grid of values of the inertial parameters.
The conclusion was made that Amalthea could not reside in
$\alpha$-resonance.

Let us introduce the parameter $\omega_0 = \sqrt{3(B-A)/C}$, which
is the frequency of small-amplitude oscillations of a satellite in
synchronous resonance (see \citep{WPM84,S99}). It roughly
characterizes the dynamical asymmetry of the satellite shape. If
we accept the data of~\citet{EA99}, then $\omega_0 = 1.058$ for
Prometheus and $\omega_0 = 0.812$ for Pandora.

For two centers of synchronous resonance to coexist in the phase
space of rotational motion (the ``Amalthea effect''), the
$\omega_0$ parameter must exceed unity slightly; see Figs. 1 and 2
in \citep{MS00} and also Fig.~\ref{fig:3} in the present paper.
Note that, in addition, the orbital eccentricity must obey a
certain constraint: it must not be too high. The limit on the
eccentricity, which allows coexistence of the periodic solutions
corresponding to $\alpha$-resonance and $\beta$-resonance, is as
follows: $e < \frac{4 \sqrt{3}}{9} (\omega_0 - 1)^{3/2}$
(\citet[Ch.~2]{B65}; \citet[p.~366]{M90}).

The ``Amalthea effect'' takes place for prolate satellites.
Indeed, $\omega_0 = \sqrt{3(B-A)/C} = \sqrt{3(a^2-b^2)/(a^2+b^2)}$
for a triaxial ellipsoid with homogeneous density; see
\cite[p.~396]{KS06}. The parameter $\omega_0
> 1$ if $c < b < a/\sqrt{2}$. In other words, two semiaxes should
be less than $\approx 0.7$ of the third one. The ``Amalthea
effect'' was considered and discussed in detail
in~\citep{KS06,MS07}. It was shown that this effect might be
abundant amongst minor planetary satellites (the satellites with
diameters less than 100 km) moving in close-to-circular orbits.

\subsection{Synchronous states: $\alpha$-resonance, $\beta$-resonance
and period-doubling bifurcation mode of $\alpha$-resonance}
\label{sec:ab_res}

The available data on the parameters of the figures of Prometheus
and Pandora are collected in Table~\ref{tab:1}. Note that the data
on the shapes of Prometheus and Pandora derived by \cite{T89} are
tabulated in the reports by \cite{SAB02}, \cite{SAA05}, and
\cite{SAA07}. Besides, the data on the adopted positions of the
poles of Prometheus and Pandora can be found in these reports;
these data are the same in the three references. We set the
orbital eccentricity equal to $e = 0.002$ for Prometheus and $e =
0.004$ for Pandora, as in \citet{EA99}. According to
\citet[Figs.~5 and 6]{GR03a}, on the timescale of 20 years the
eccentricity of Prometheus varies in the limits
$2.27\times10^{-3}$~-- $2.30\times10^{-3}$, and that of Pandora in
the limits $4.35\times10^{-3}$~-- $4.38\times10^{-3}$; see also
\citep{F03}.

\begin{table}
\caption{The shape parameters for Prometheus and Pandora}
\label{tab:1}
\begin{tabular}{lllll}
\hline\noalign{\smallskip}
\multicolumn{2}{c}{Prometheus (S16)} &
\multicolumn{2}{c}{Pandora (S17)} & References \\
$b/a$ & $c/b$ & $b/a$ & $c/b$ & \\
\hline\noalign{\smallskip}
0.714 & 0.740 & 0.764 & 0.786 & \cite{W87}\\
0.676 & 0.680 & 0.800 & 0.705 & \cite{T89}\\
0.586 & 0.706 & 0.737 & 0.738 & \cite{S93}\\
0.608 & 0.726 & 0.741 & 0.782 & \cite{GM95}\\
0.676 & 0.680 & 0.800 & 0.705 & \cite{EA99}\\
0.734 & 0.696 & 0.773 & 0.804 & \cite{P06}\\
\noalign{\smallskip}\hline
\end{tabular}
\end{table}

\begin{figure}
\begin{tabular}{c}
{\bf a)} \includegraphics[width=0.75\textwidth]{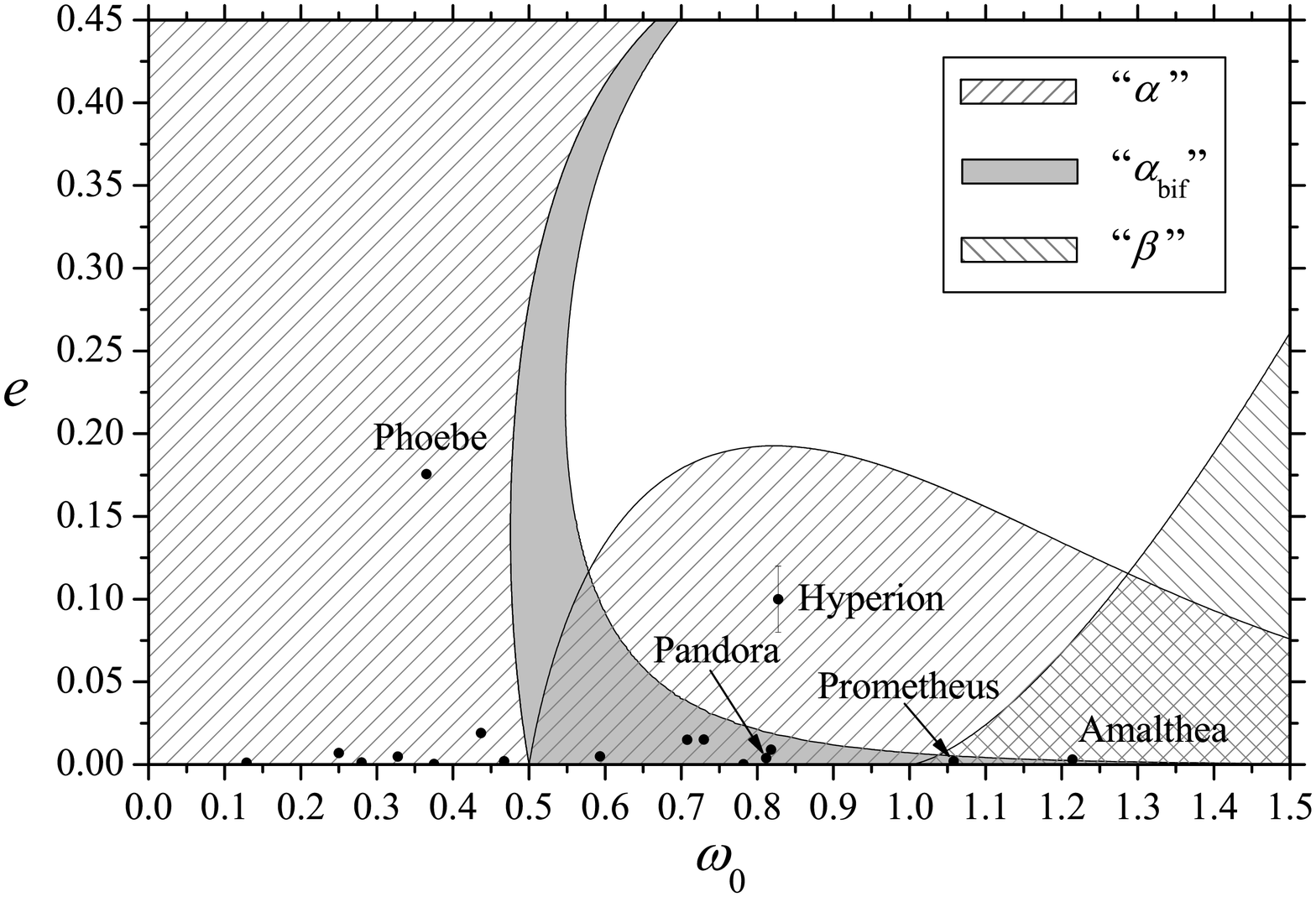}\\
{\bf b)} \includegraphics[width=0.75\textwidth]{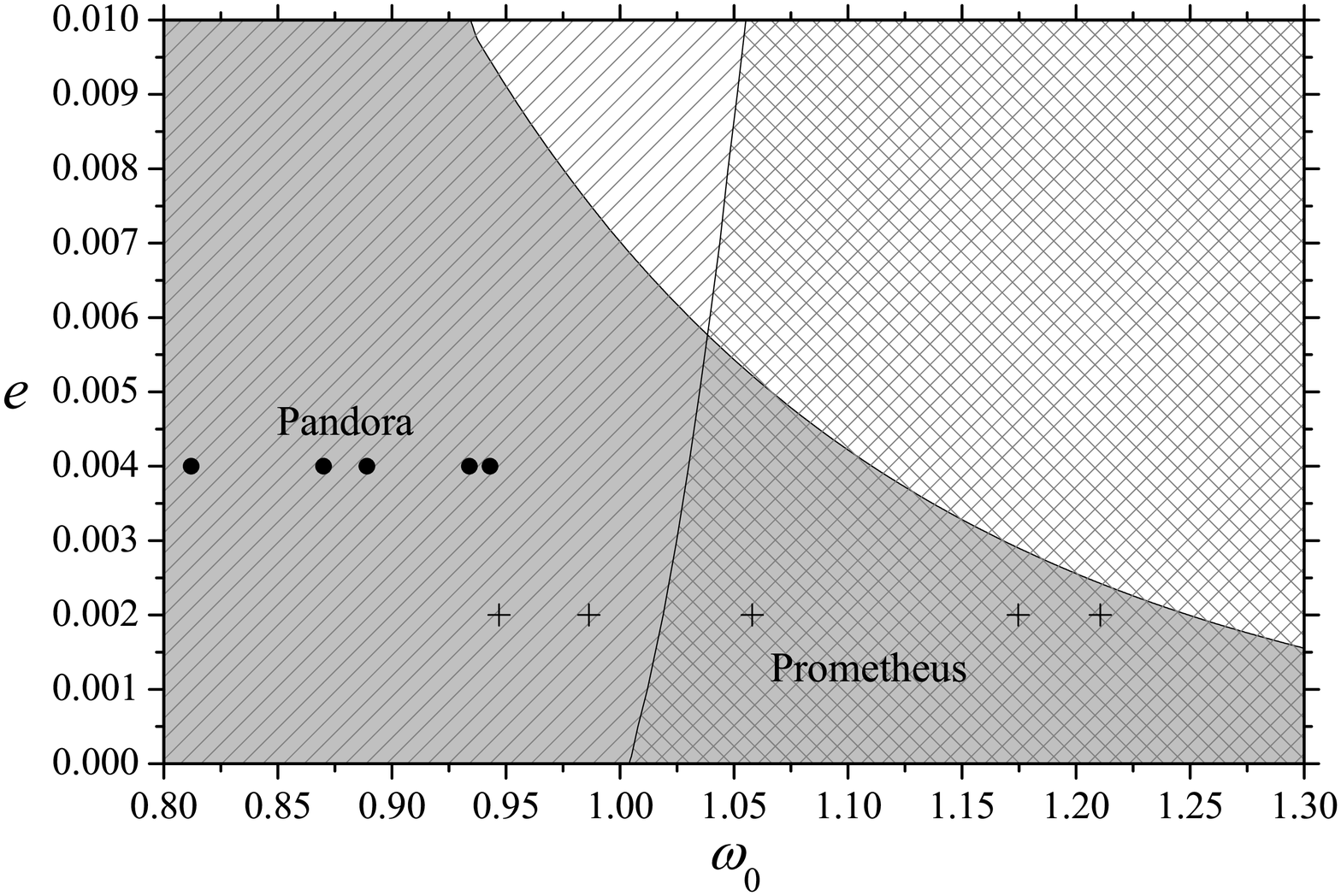}\\
\end{tabular}
\caption {{\bf (a)} Location of the satellites with known values
of the shape parameters in the ``$e$~--- $\omega_0$'' diagram. {\bf
(b)} Location of Prometheus (crosses) and Pandora (circles) in the
diagram ``$e$~--- $\omega_0$'' according to Table~\ref{tab:1}}
\label{fig:1}
\end{figure}

The ``$e$~--- $\omega_0$'' diagram for the satellites with known
values of the shape parameters, based on the data compiled
in~\citep{KS05}, is shown in Fig.~\ref{fig:1}(a). The theoretical
boundaries of the zones of existence of $\alpha$-resonance,
$\beta$-resonance, and period-doubling bifurcation mode
$\alpha_\mathrm{bif}$ of $\alpha$-resonance are indicated
according to the data in~\citep{M01}.

As follows from Fig.~\ref{fig:1}(a), period-doubling bifurcation
mode $\alpha_\mathrm{bif}$ of $\alpha$-resonance can be present in
the phase space of rotation of Prometheus and Pandora. For
Pandora, this confirms an earlier analysis by \cite{M01} made on
the basis of a single estimate of the inertial parameters. In
Fig.~\ref{fig:1}(b), a part of the ``$e$~--- $\omega_0$'' diagram
with location of Prometheus and Pandora indicated according to the
data of Table~\ref{tab:1} is given in higher resolution. In the
case of Pandora, $\beta$-resonance does not exist for all the data
considered. In the case of Prometheus, $\beta$-resonance does not
exist only for the data due to~\citet{W87} and~\citet{P06}.
According to all other sources, Prometheus lies in the zone of
existence of this mode. For all the data on Prometheus and
Pandora, period-doubling bifurcation mode of $\alpha$-resonance
exists.

\begin{figure}
\includegraphics[width=0.75\textwidth]{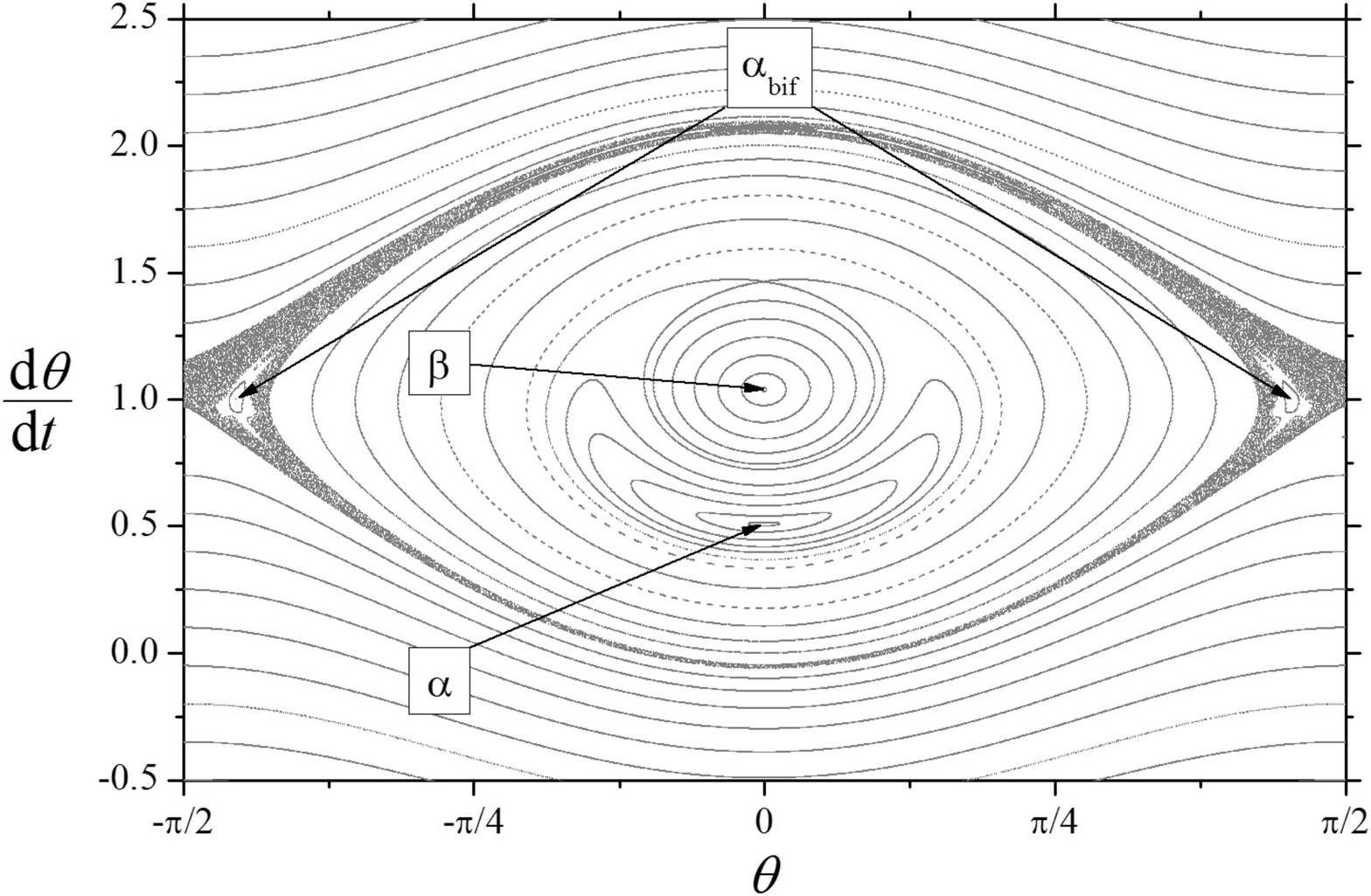}
\caption{The phase space section of the planar rotational motion
of Prometheus. The exact synchronous resonance modes are
indicated.} \label{fig:2}
\end{figure}

In Fig.~\ref{fig:2}, the phase space section of the planar
rotational motion of Prometheus is shown. The section is defined
at the pericenter of the orbit; i.e., the variables are mapped
each orbital period. Note that the planar problem (that with $\phi
= \psi = 0$) has one and a half degrees of freedom. The center of
$\alpha$-resonance (the lower one in the section) and that of
$\beta$-resonance (the upper one) are indicated in the section.
Besides, there is period-doubling bifurcation mode
$\alpha_\mathrm{bif}$, that manifests itself in the two prominent
regular islands inside the chaotic layer at the left and at the
right sides from the center of $\beta$-resonance.

We see that Prometheus is subject to the ``Amalthea effect''. In a
different terminology this was noted in \citep{M01}. What is more,
period-doubling bifurcation mode of $\alpha$-resonance is possible
for this satellite, i.e., the potentially possible rotational
dynamics are very rich.

\subsection{Attitude stability of exact synchronous rotation}
\label{sec:stab_syn}

Let us consider the stability of synchronous rotation of
Prometheus and Pandora with respect to tilting the axis of
rotation in the cases of exact $\alpha$-resonance,
$\beta$-resonance, and period-doubling bifurcation mode
$\alpha_\mathrm{bif}$. \cite{MS00} and \cite{KS05} performed a
research of the stability on the ($A/C$, $B/C$) plane. Here we use
another plane, namely, the ($c/b$, $b/a$) plane\footnote{We are
grateful to A.\,Dobrovolskis for the advice to use the given
coordinates.}, which is more graphical.

In the cases of $\alpha$-resonance and $\beta$-resonance we use a
method based on the analysis of the modal structure of the
differential distribution of the computed modules of multipliers
of periodic solutions of the equations of motion. The method is
described in detail in~\citep{MS00}.

In the considered problem, the system of equations of motion in
variations with respect to the periodic solution consists of six
linear differential equations of the first order with periodic
coefficients. Numerical integration of the system allows one to
obtain the matrix of linear transformation of variations for one
period; see \citep{WPM84}. The periodic solutions in the given
problem are characterized by three pairs of multipliers. The
distributions of the modules of multipliers are built for a set of
trajectories corresponding to a center of synchronous resonance on
a grid of values of the $b/a$ and $c/b$ parameters. Analysis of
the distributions (see \cite{MS00}) allows one to separate orbits
stable with respect to tilting the axis of rotation from those
which are unstable.

Analysis of the attitude stability of period-doubling bifurcation
mode $\alpha_\mathrm{bif}$ is carried out by means of computation
of the whole spectrum of the Lyapunov characteristic exponents
(LCEs) for a set of values of the $b/a$ and $c/b$ parameters,
followed by analysis of the differential distributions of the
computed values of the LCEs. By means of a similar method we
investigated the stability of rotation of planetary
satellites~\citep{MS98} on sets of initial data of the
trajectories. Here the distributions of the LCE values for the
trajectories with the initial data taken at exact period-doubling
bifurcation mode $\alpha_\mathrm{bif}$ are built on a grid of
values of the $b/a$ and $c/b$ parameters. We set the grid
resolution equal to $0.001$ in both $b/a$ and $c/b$ axes. Analysis
of the modal structure of the distributions allows one to separate
the stable trajectories, for which all three indices are zero, and
the unstable ones for which at least one of the LCEs is distinct
from zero.

The true values of the LCEs are supposed to be the limits of the
computed values when the time of computation tends to infinity.
The time of computation is necessarily finite. However, it is
implied henceforth that the obtained numerical values in the case
of chaotic trajectories represent the true LCE values, because the
time of computation was taken to be long enough for the computed
LCEs of the chaotic trajectories to saturate (i.e., increasing the
computation time would not make the computed LCE values less; a
``plateau'' is reached in each case). In what concerns the regular
trajectories, the obtained numerical values of the LCEs tend to
zero with increasing the computation time.

For the computation of the LCE spectra, we use the HQRB
method~\citep{BUP97}, programmed as a software package
in~\citep{SK02, KS03}. The Dormand--Prince integrator
DOPRI8~\citep{HNW87}, realizing in Fortran the 8th order
Runge--Kutta method with the step size control, is used for
integration of equations of motion~(\ref{eyleq}),
(\ref{angvel_1}).

\begin{figure}
\includegraphics[width=0.5\textwidth]{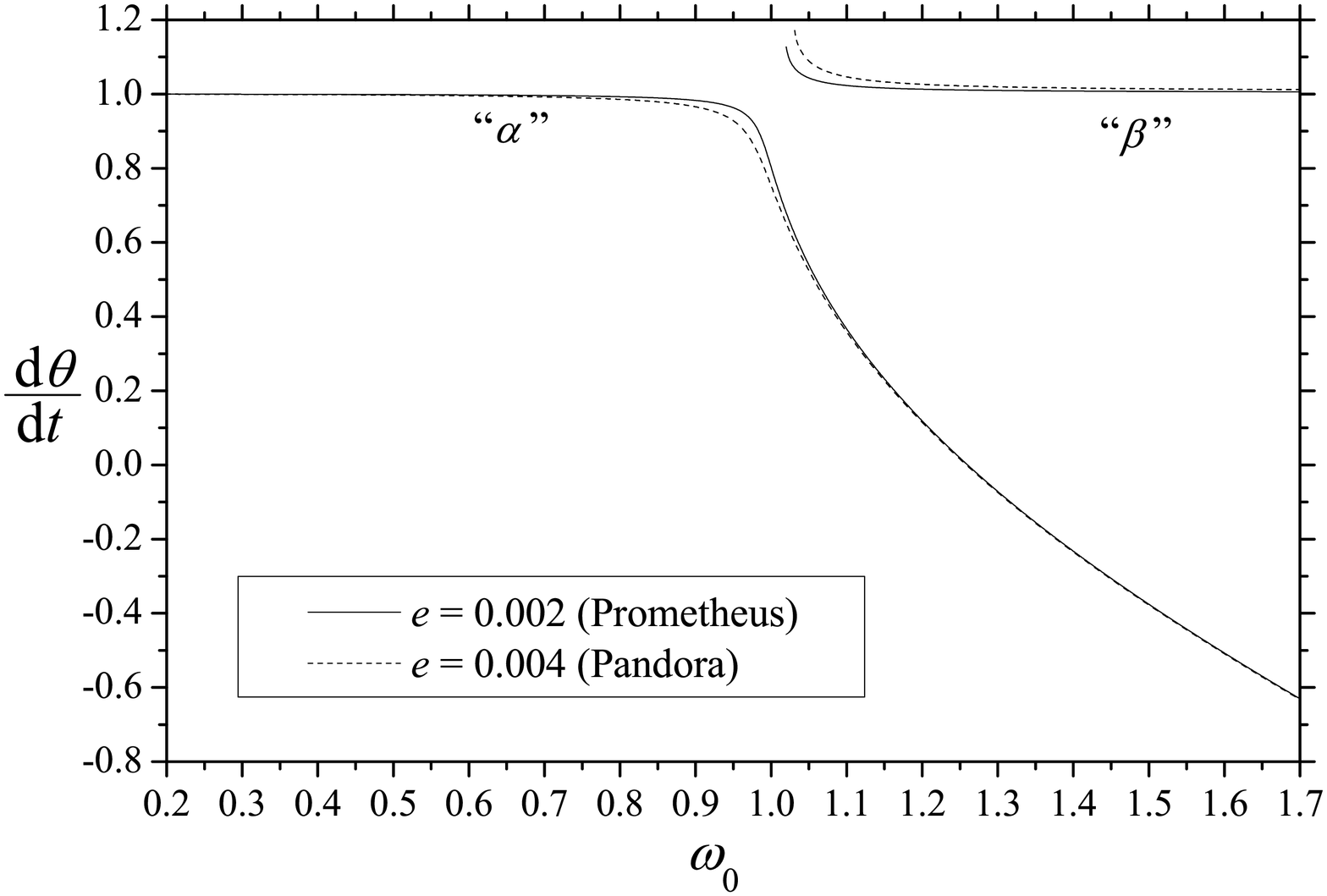}
\caption{The $\mathrm{d}\theta/\mathrm{d}t$ coordinates of the
centers of synchronous resonance in the phase space section in
dependence on the $\omega_0$ parameter. The curves starting on the
left side of the plot correspond to $\alpha$-resonance, those
starting on the right side correspond to $\beta$-resonance.}
\label{fig:3}
\end{figure}

\begin{figure}
\begin{tabular}{c}
{\bf a)} \includegraphics[width=0.5\textwidth]{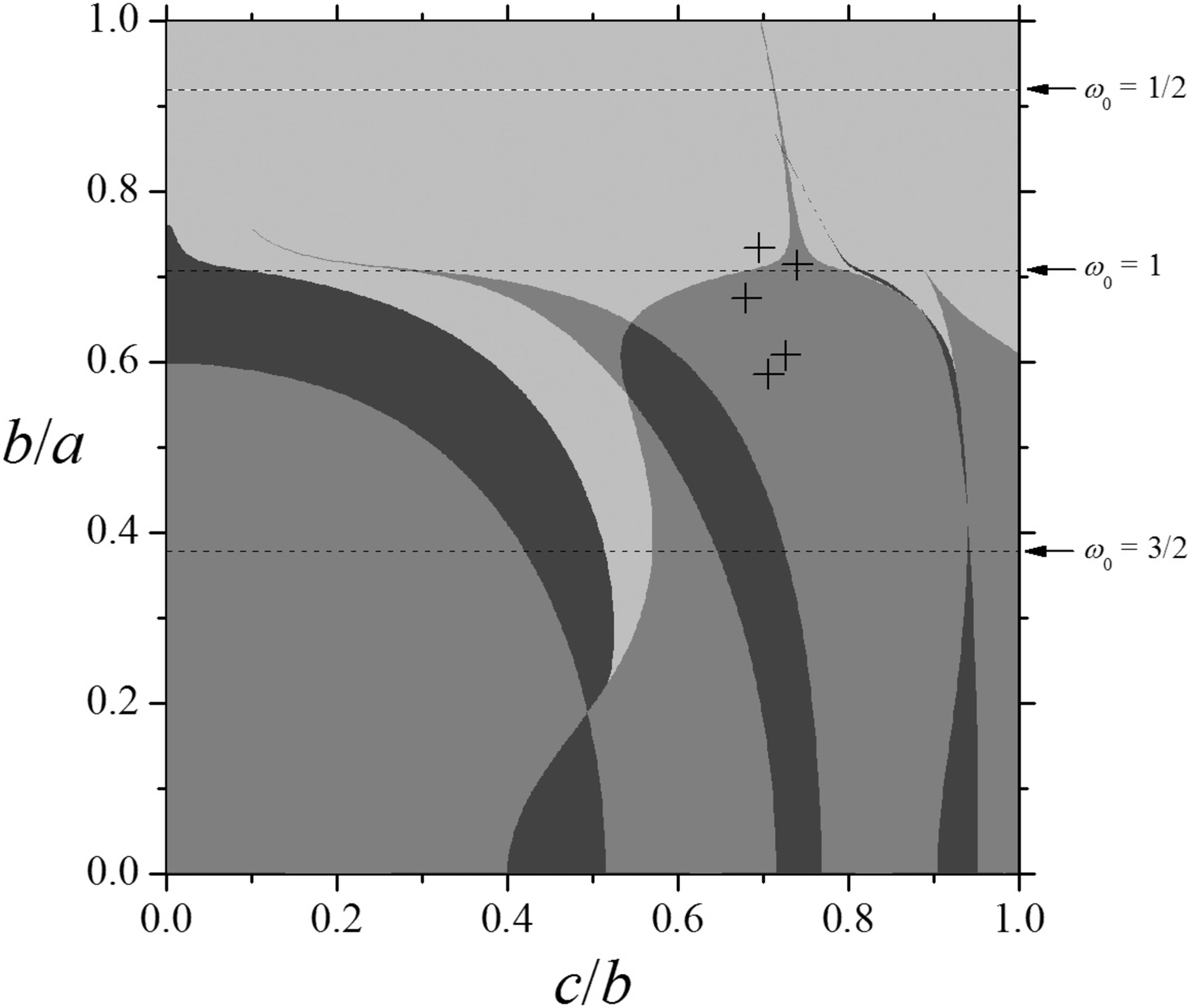}\\
{\bf b)} \includegraphics[width=0.5\textwidth]{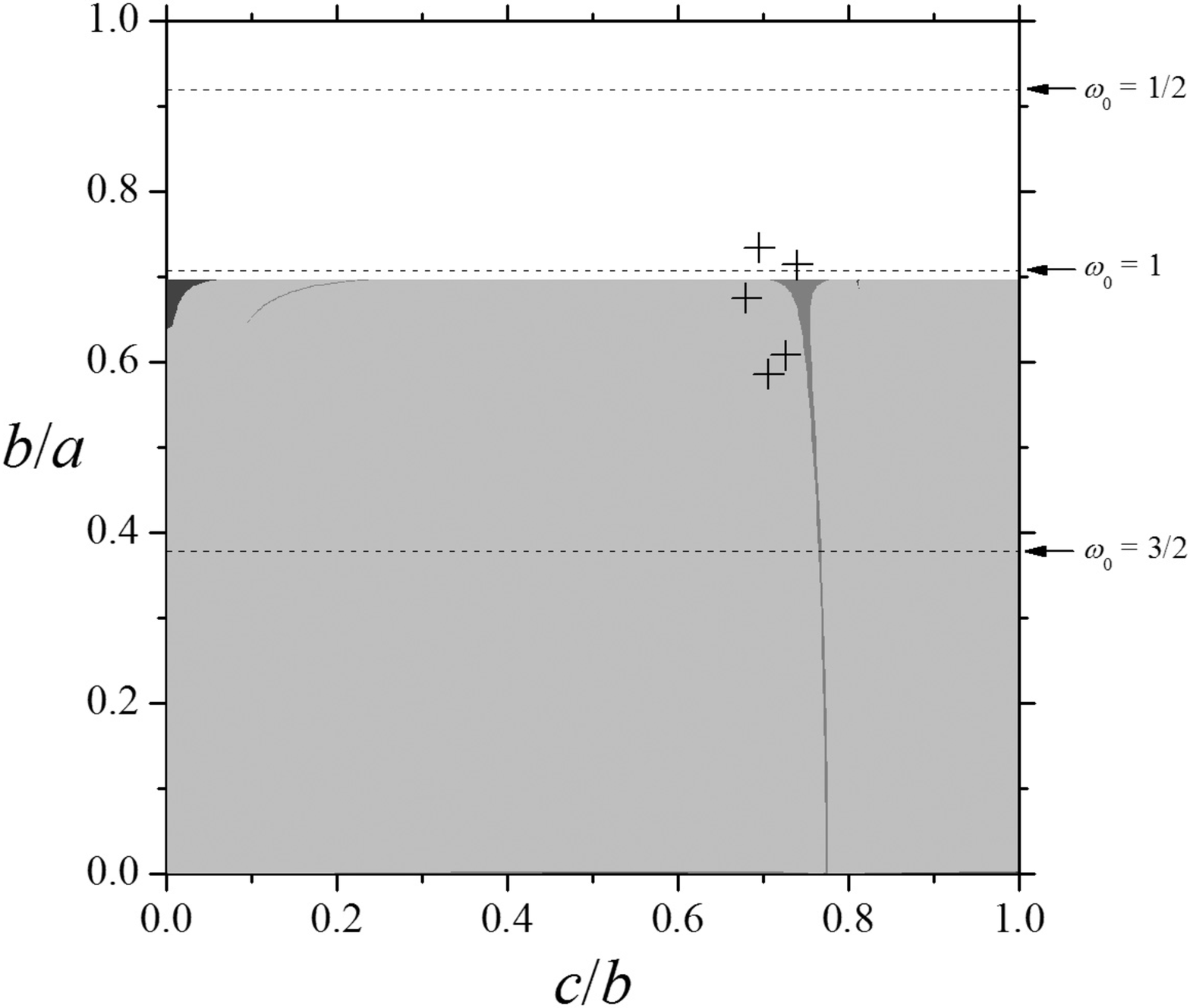}\\
{\bf c)} \includegraphics[width=0.5\textwidth]{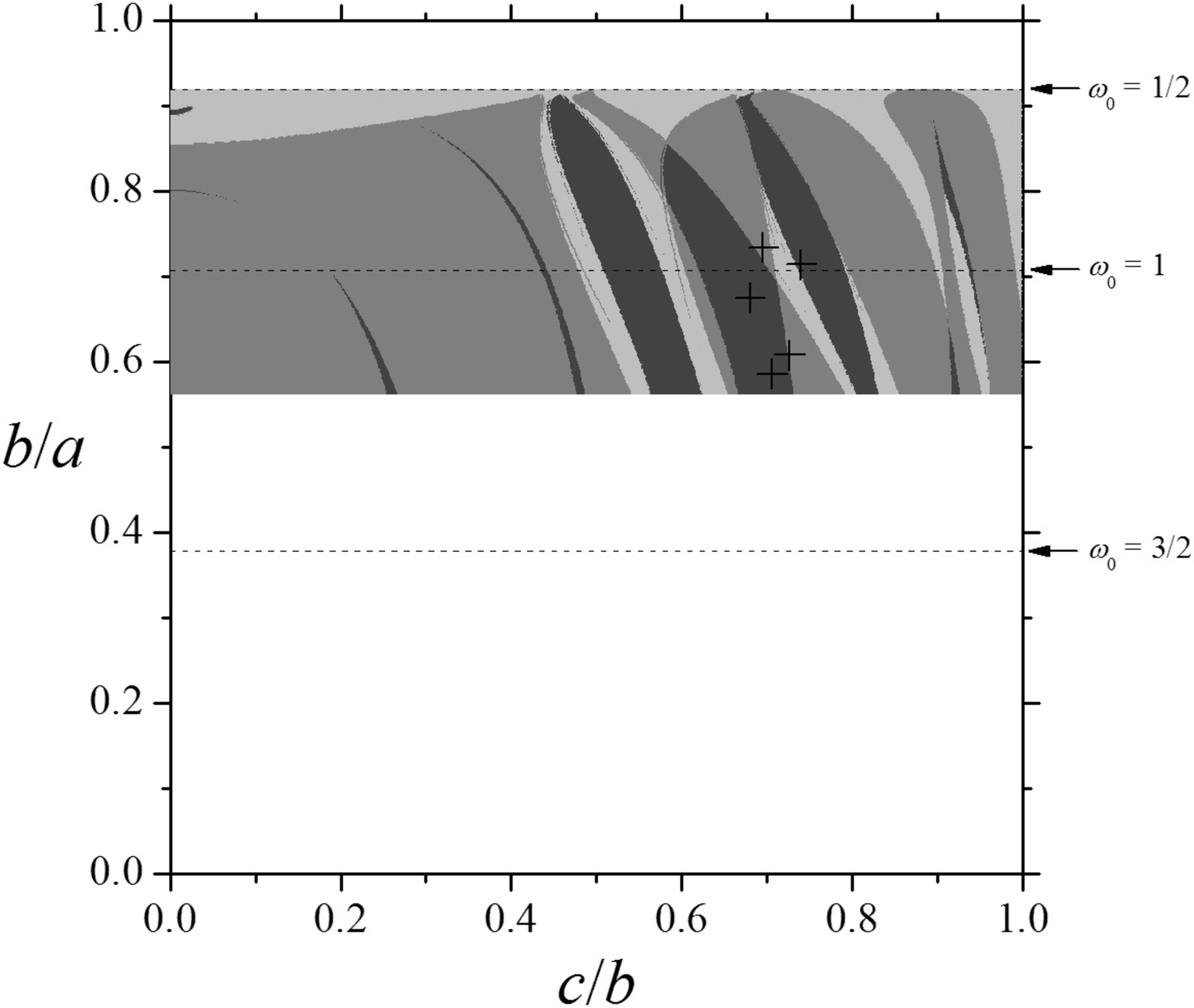}\\
\end{tabular}
\caption{Regions of stability and instability with respect to
tilting the axis of rotation; $e = 0.002$ (Prometheus): {\bf (a)}
for the center of $\alpha$-resonance, {\bf (b)} for the center of
$\beta$-resonance, {\bf (c)} for the exact period-doubling
bifurcation mode of $\alpha$-resonance. The locations of
Prometheus according to the data of Table~\ref{tab:1} are
indicated by crosses.} \label{fig:4}
\end{figure}

In Fig.~\ref{fig:3}, the $\mathrm{d}\theta/\mathrm{d}t$
coordinates of the centers of synchronous resonance in the phase
space section in dependence on the $\omega_0$ parameter are shown
for the values of the orbital eccentricities of both satellites.
The curves starting on the left side of the plot correspond to
$\alpha$-resonance, those starting on the right side correspond to
$\beta$-resonance. As it is clear from Fig.~\ref{fig:3}, $\alpha$-
and $\beta$-resonances coexist in a substantial interval of
$\omega_0$, if the orbital eccentricities are so small.
$\beta$-resonance is born at $\omega_0 \approx 1$;
$\alpha$-resonance disappears at a large value of $\omega_0$, out
of the presented plot limit. The problem of bifurcations causing
the birth and disappearance of $\alpha$- and $\beta$-resonances is
considered in brief by \cite{MS00}.

\begin{figure}
\begin{tabular}{c}
{\bf a)} \includegraphics[width=0.5\textwidth]{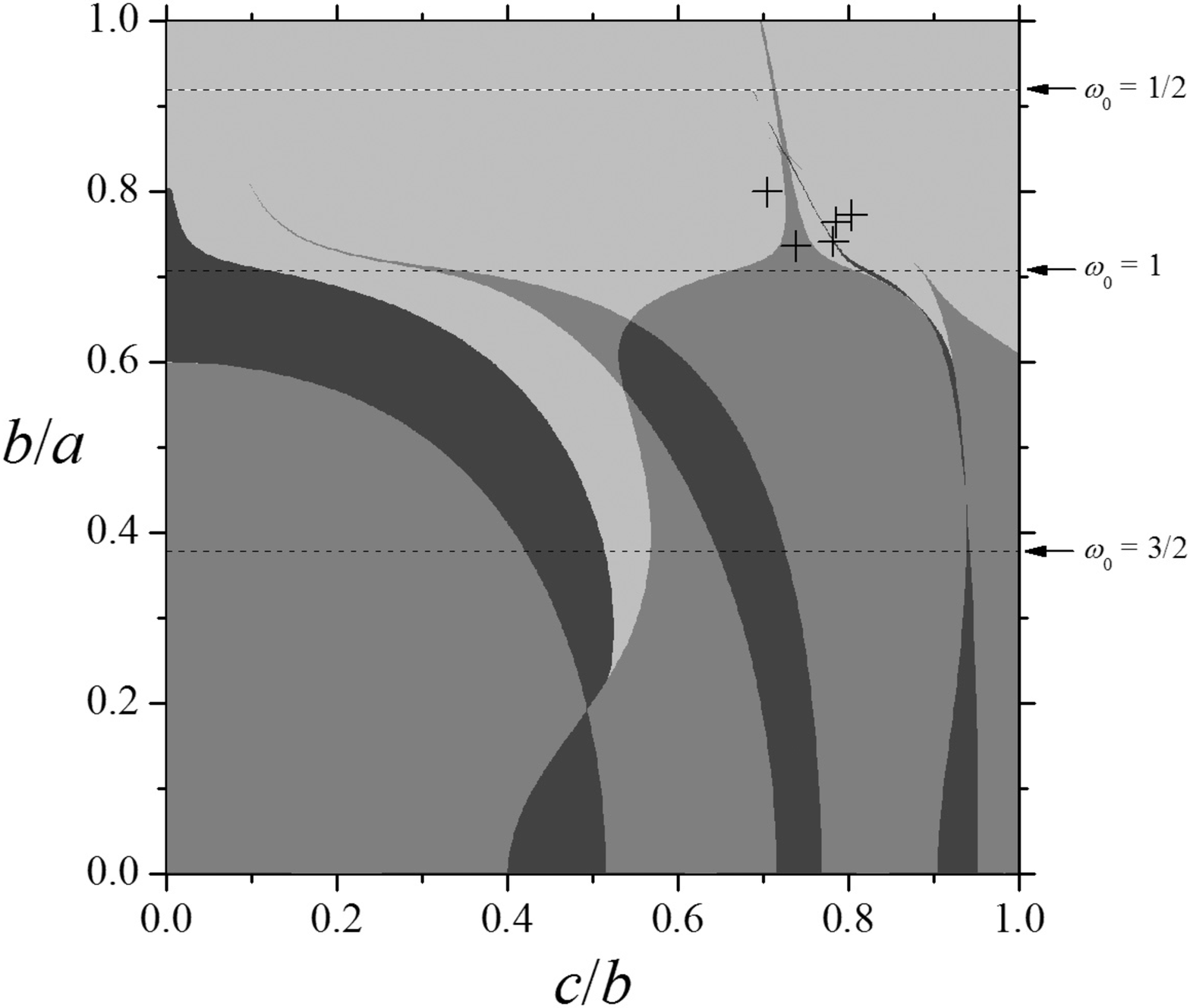}\\
{\bf b)} \includegraphics[width=0.5\textwidth]{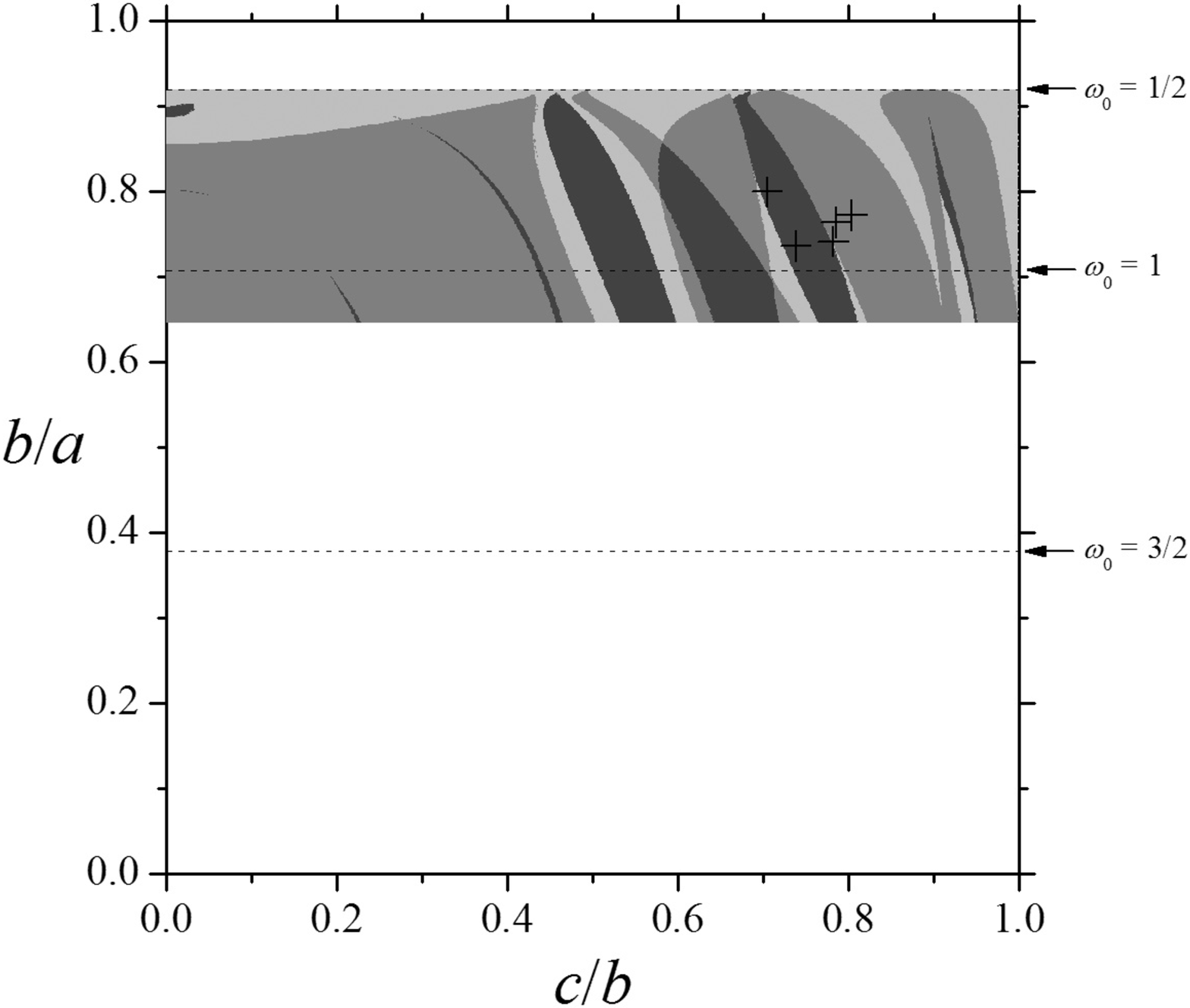}\\
\end{tabular}
\caption{Regions of stability and instability with respect to
tilting the axis of rotation; $e = 0.004$ (Pandora): {\bf (a)} for
the center of $\alpha$-resonance, {\bf (b)} for the exact
period-doubling bifurcation mode of $\alpha$-resonance. The
locations of Pandora according to the data of Table~\ref{tab:1}
are indicated by crosses.} \label{fig:5}
\end{figure}

The computed regions of stability and instability are shown in
Fig.~\ref{fig:4} (for Prometheus) and in Fig.~\ref{fig:5} (for
Pandora). The regions of stability are shown in light gray, the
regions of minimum (one degree of freedom) instability are shown
in dark gray, and the regions of maximum (two degrees of freedom)
instability are shown in black. The lines of constant value of the
$\omega_0$ parameter are depicted for orientation.

It follows from the diagrams in Figs.~\ref{fig:4}(a) and (b) that
$\alpha$-resonance is attitude unstable for most of the
observational data for Prometheus. Earlier this instability was
noted in \citep{M01,KS05} on the basis of a single estimate of the
inertial parameters. In the case of the data due to~\cite{P06},
$\alpha$-resonance is close to instability. Rotation of Prometheus
in $\beta$-resonance is close to instability, in agreement with an
inference by \cite{KS05}. In the case of Pandora (see
Fig.~\ref{fig:5}(a)), rotation in $\alpha$-resonance is close to
instability for the data of~\citep{W87, T89, P06} and is unstable
for the data of~\citep{S93, GM95}, in general agreement with
\citep{KS05}, where the stability analysis was based on a single
estimate of the inertial parameters.

From the diagrams in Fig.~\ref{fig:4}(c) and Fig.~\ref{fig:5}(b)
it is clear that synchronous rotation in the $\alpha_\mathrm{bif}$
mode is attitude unstable for both Prometheus and Pandora.

\subsection{Attitude stability in the general case}
\label{sec:arb_rot}

Not only the stability in exact synchronous rotation is of
interest. Let us study the attitude stability of the trajectories
in the vicinities of the exact synchronous states on
representative sets of initial data. Computation of the LCEs on a
grid of initial data, followed by analysis of the distributions of
the computed LCEc, enables one to accomplish such a study. A
similar study with the use of the maximum LCEs was carried out
in~\citep{MS98} for Phobos, Deimos, Amalthea, and Hyperion. Here
we study Prometheus and Pandora by means of analysis of the LCE
spectra. Besides, we significantly increase the resolution of the
initial data grid, as well as the time interval on which the LCEs
are computed. For the surface of section we choose the ($\theta$,
$\mathrm{d}\theta/\mathrm{d}t)$ plane taken at $t = 2 \pi m$, $m =
0$, 1, 2, $\ldots$, i.e., defined at the orbit pericenter. In the
computations we adopt the values of $b/a$ and $c/b$ as given
in~\citet{EA99} (see Table~1).

The computation of the LCEs has been carried out for two sets of
trajectories, (i) and (ii), defined by the following choice of
initial data: (i) $\theta = 0$, (ii) $\theta = \pi/2$; and in the
both sets $\mathrm{d}\theta/\mathrm{d}t$ is taken in the range
from $-0.5$ to $2.5$ with the step equal to $0.003$ in the case of
Prometheus, and in the range from $0.0$ to $2.0$ with the step
equal to $0.002$ in the case of Pandora. The initial conditions
also include $\phi = \psi = 0$, $\mathrm{d}\phi/\mathrm{d}t =
\mathrm{d}\psi/\mathrm{d}t = 0$; the motion starts (i.e., $t = 0$)
at the pericenter. Thus the LCEs have been computed for 2000
trajectories for each satellite.

Following~\cite{MS98}, in order to separate the regular and
chaotic orbits, we build differential distributions of the
computed LCE values. The distribution has two peaks; one of them
corresponds to the chaotic orbits, and one to the regular orbits.
On increasing the time interval of integration, the peak
corresponding to the chaotic orbits remains motionless, while the
abscissa of the peak corresponding to the regular orbits tends to
zero (or to minus infinity in the logarithmic scale). Thus the
sets of regular and chaotic trajectories are separated. The
abscissa of a point between the peaks gives the numeric criterion
for separation of the sets. Increasing the time interval of
integration allows one to make this numeric criterion more
precise.

\begin{figure}
\begin{tabular}{c}
{\bf a)} \includegraphics[width=0.75\textwidth]{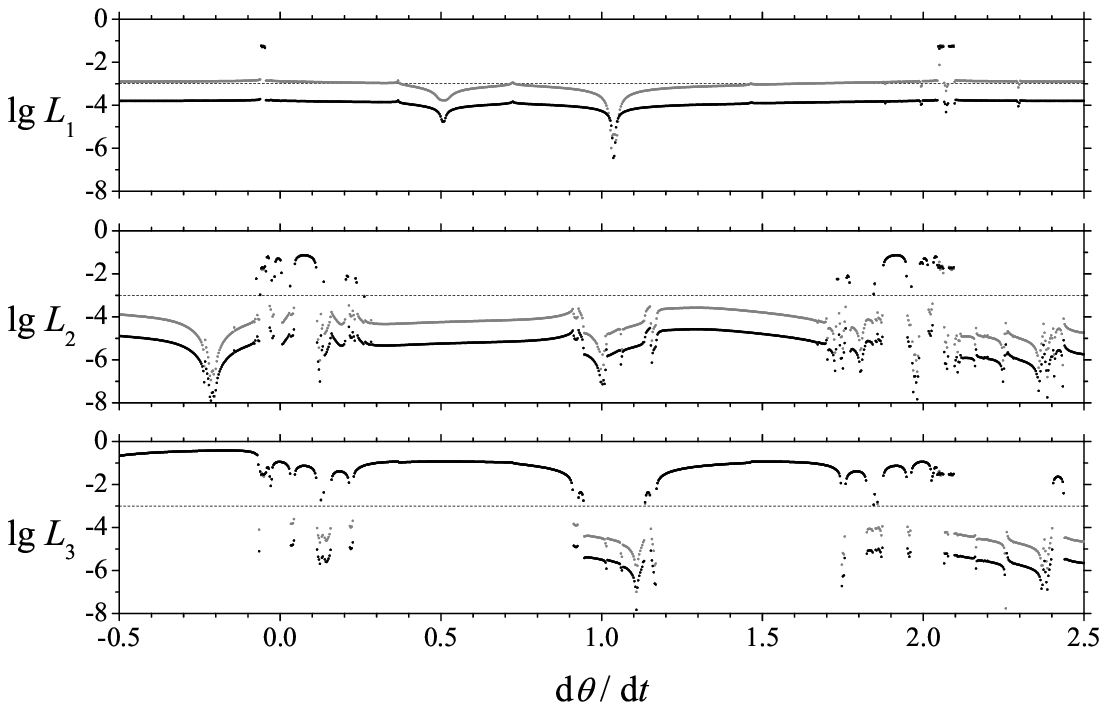}\\
{\bf b)} \includegraphics[width=0.75\textwidth]{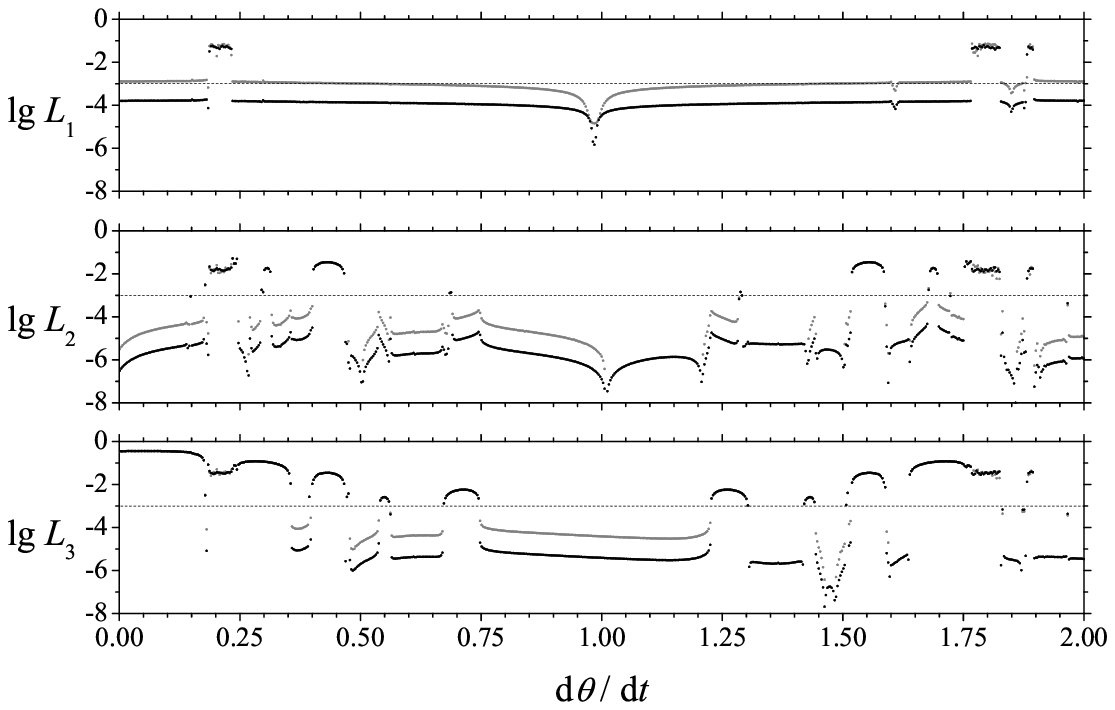}\\
\end{tabular}
\caption{The dependence of the LCEs on initial data. {\bf (a)} $e
= 0.002$, $\omega_0 = 1.058$ (Prometheus); {\bf (b)} $e = 0.004$,
$\omega_0 = 0.812$ (Pandora). The gray curves correspond to the
computation time $t = 10^4$, the black curves to $t = 10^5$. The
horizontal dashed lines correspond to the adopted value of the LCE
decimal logarithm (equal to $-3$ in all cases) separating the
chaotic and regular trajectories.} \label{fig:6}
\end{figure}

The LCE dependences on the initial value of
$\mathrm{d}\theta/\mathrm{d}t$ in the case of $\theta = 0$ are
shown in Fig.~\ref{fig:6}. They have been computed on the time
intervals $t = 10^4$ and $10^5$. From the plots it is clear that
with increasing the integration time the computed LCE values for
the chaotic trajectories remain constant, while the computed LCE
values for the regular trajectories decrease. The horizontal
dashed lines in the plots correspond to the adopted value of the
LCE decimal logarithm (equal to $-3$ in all cases) separating the
chaotic and regular trajectories. This criterion has been derived
by means of building the distributions of the computed values of
the LCEs at $t = 10^5$. It separates the peaks corresponding to
the chaotic and regular trajectories in the distributions.

\begin{figure}
\begin{tabular}{c}
{\bf a)} \includegraphics[width=0.75\textwidth]{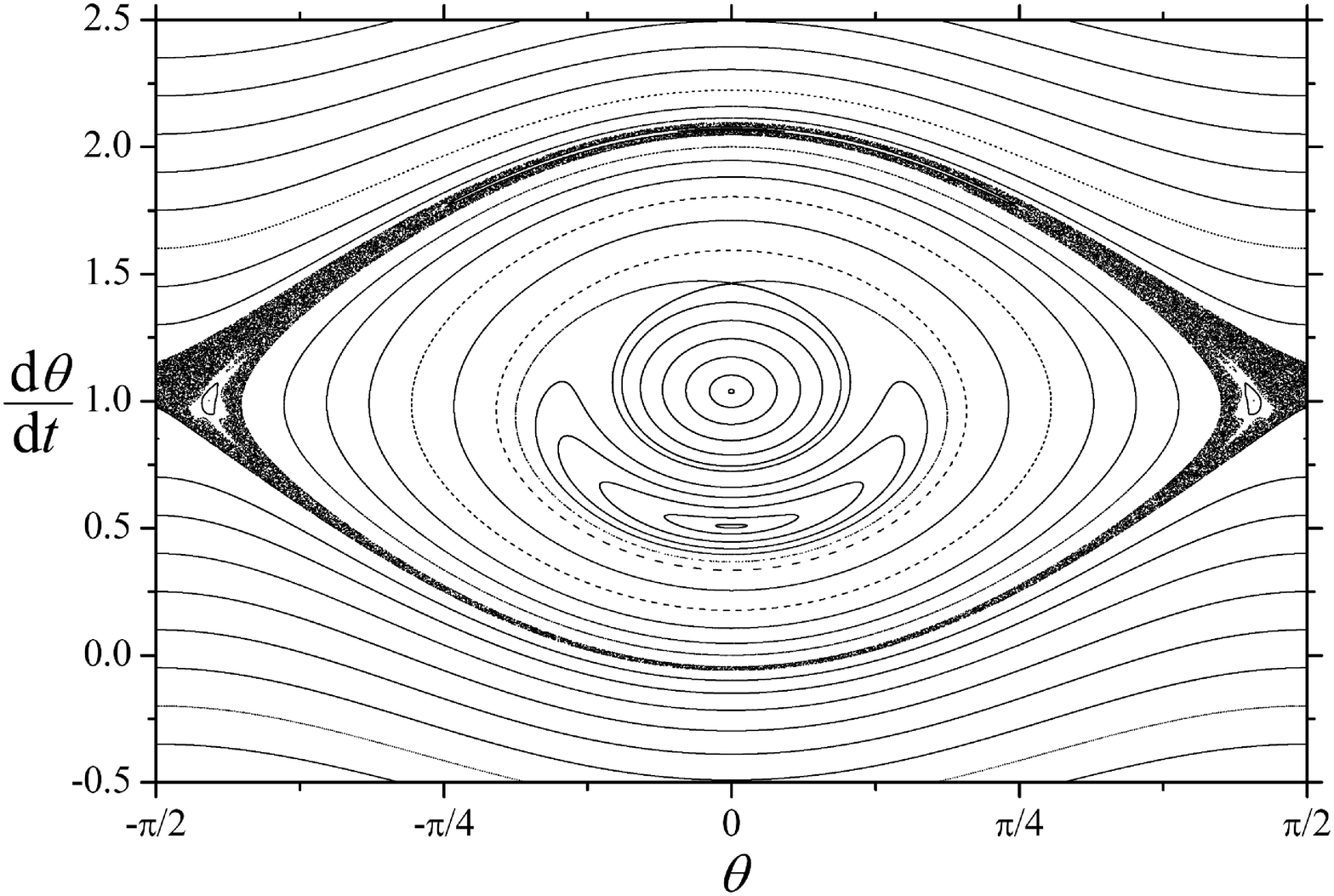}\\
{\bf b)} \includegraphics[width=0.75\textwidth]{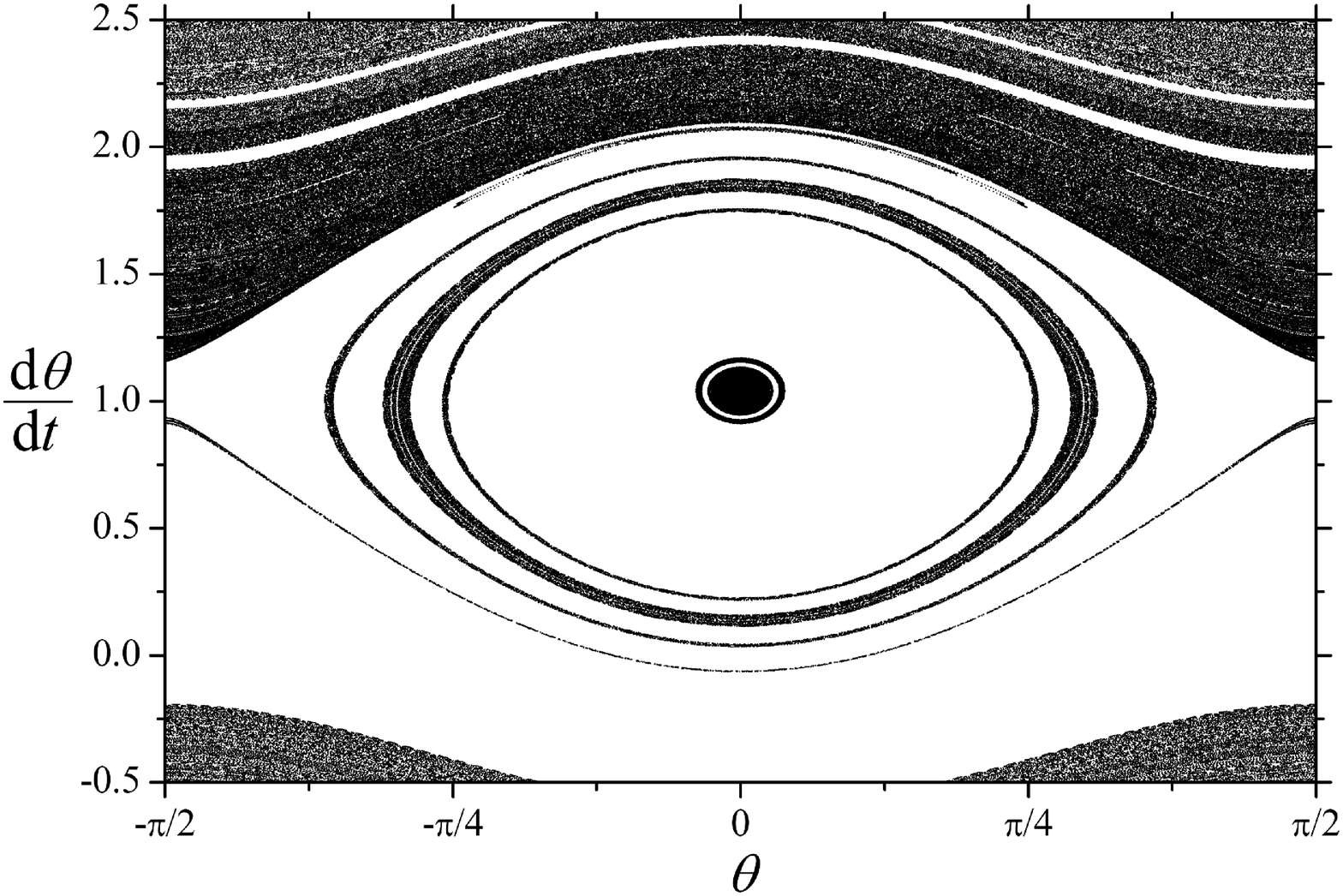}\\
\end{tabular}
\caption{The phase space section for $e = 0.002$, $\omega_0 =
1.058$ (Prometheus): {\bf (a)} all trajectories, {\bf (b)} only
attitude stable ones.} \label{fig:7}
\end{figure}

\begin{figure}
\begin{tabular}{c}
{\bf a)} \includegraphics[width=0.75\textwidth]{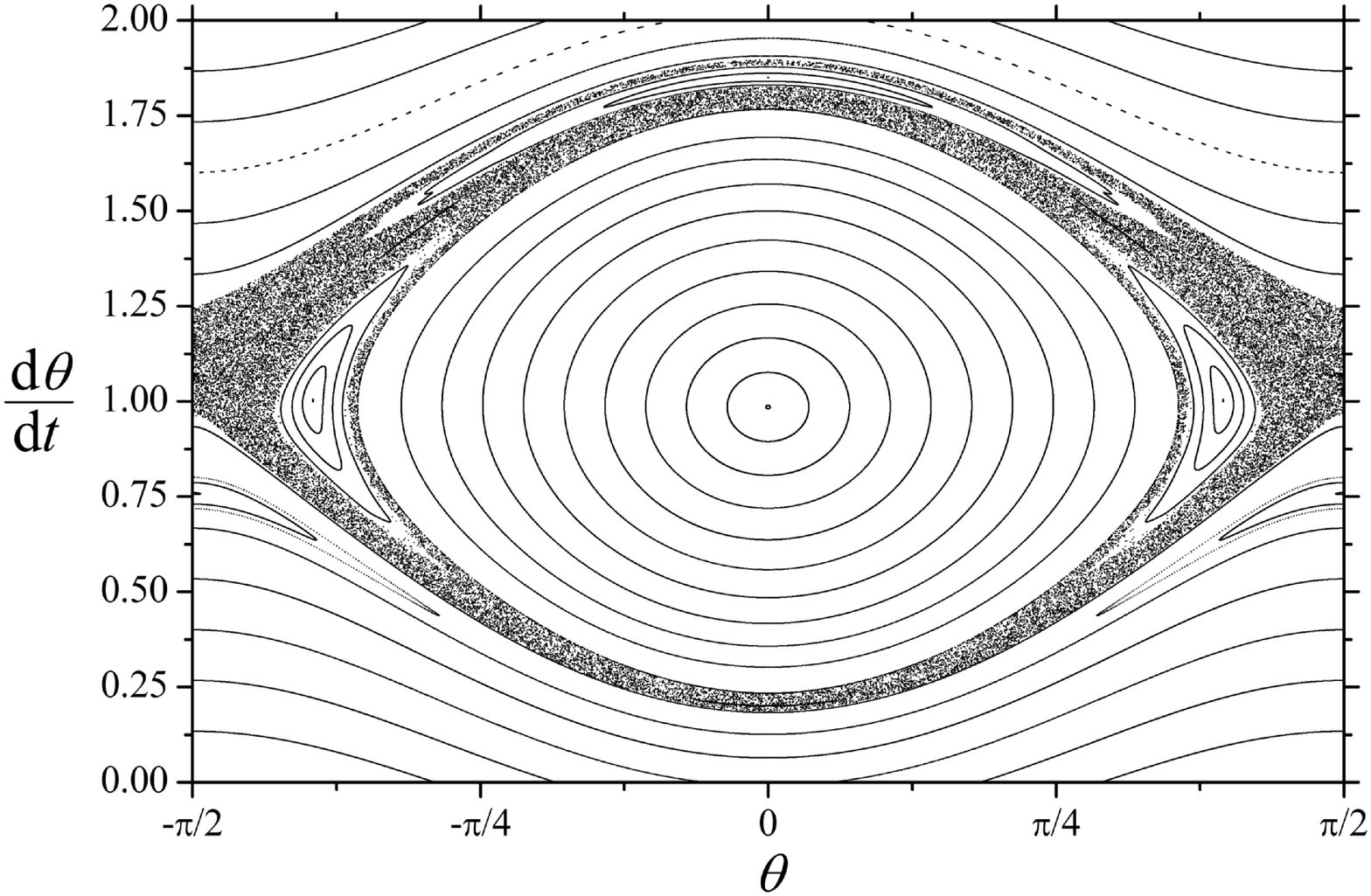}\\
{\bf b)} \includegraphics[width=0.75\textwidth]{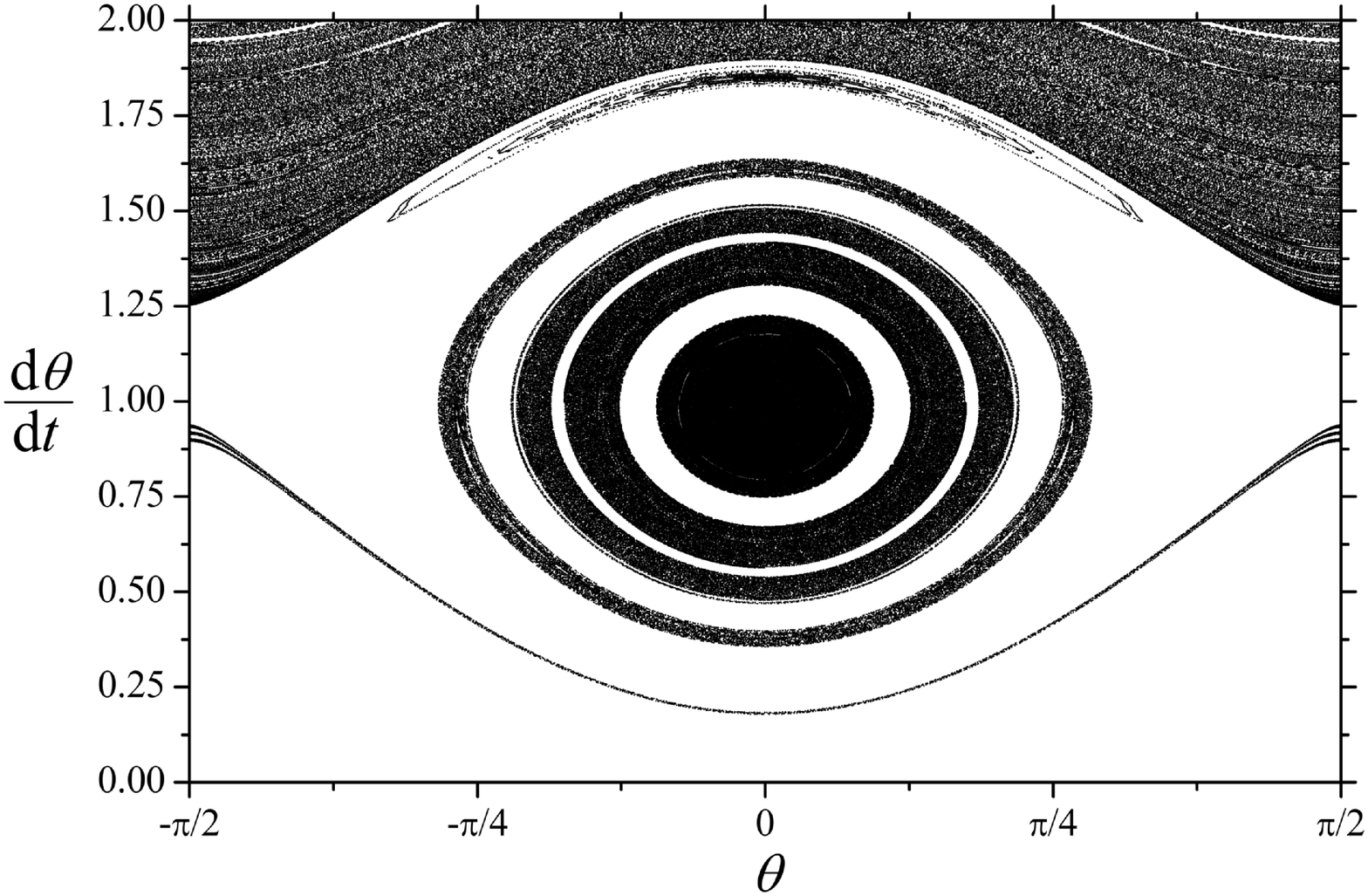}\\
\end{tabular}
\caption{The phase space section for $e = 0.004$, $\omega_0 =
0.812$ (Pandora): {\bf (a)} all trajectories, {\bf (b)} only
attitude stable ones.} \label{fig:8}
\end{figure}

The separation of the regular and chaotic orbits, accomplished by
means of analysis of the LCE distributions, allows one to build
two variants of the phase space section, one for all trajectories,
and the other one only for the attitude stable trajectories. These
double variants are shown for Prometheus in Figs.~\ref{fig:7}(a)
and \ref{fig:7}(b), and for Pandora in Figs.~\ref{fig:8}(a) and
\ref{fig:8}(b). The phase space sections for all trajectories are
shown in Figs.~\ref{fig:7}(a) and \ref{fig:8}(a), while those for
the attitude stable trajectories solely are shown in
Figs.~\ref{fig:7}(b) and \ref{fig:8}(b). In order that structure
of the section containing all trajectories were clearly
discernible, it is constructed with a relatively low resolution of
the initial data grid.

In the case of Prometheus (Fig.~\ref{fig:2}, Fig.~\ref{fig:7}(a)),
there are two centers of synchronous resonance: $\alpha$-resonance
(the lower one in the section) and $\beta$-resonance (the upper
one). Besides, there exists period-doubling bifurcation mode
$\alpha_\mathrm{bif}$, located inside the chaotic layer at the
left and right sides of the center of $\beta$-resonance. The
librational trajectories are attitude stable only in the nearest
neighbourhood of the center of $\beta$-resonance. Alternation of
ring-like zones of stable and unstable motion is clearly seen for
the librational trajectories enclosing the major synchronous
state. This alternation is also clearly noticeable in
Fig.~\ref{fig:6}(a), where the computed values of the LCEs are
given in function of the initial $\mathrm{d}\theta/\mathrm{d}t$
value. A broad band of quasiperiodic trajectories under the lower
branch of the basic chaotic layer is attitude unstable. On the
contrary, the motion above the upper branch of the layer is
stable.

In the case of Pandora (Fig.~\ref{fig:8}), there exists
period-doubling bifurcation mode $\alpha_\mathrm{bif}$ located
inside the chaotic layer at the left and right sides of the center
of $\alpha$-resonance. Alternation of ring-like zones of attitude
stable and unstable librational trajectories enclosing the main
synchronous state is even more pronounced than in the case of
Prometheus.

During the process of tidal capture in synchronous resonance, both
satellites inevitably cross these intermittent belts of attitude
instability. When such belts are present, attaining exact
resonance might be more difficult in comparison with the usual
situation when they are absent, because the satellite would tend
to deviate from planar rotation in these zones, due to the
attitude instability. So, one can put forward a hypothesis that
these belts might form ``barriers'' for capturing the satellites
in synchronous rotation. Some numerical-experimental as well as
theoretical work is necessary to infer whether this hypothesis is
right or not. In particular, it is necessary to compare the
timescale of developing the attitude instability in a belt with
that of crossing the belt due to tidal evolution.

\section{Preferred orientation in chaotic rotation}
\label{sec:orient}

In a theoretical research \citep{KS05} it was found that
Prometheus and Pandora are likely to be in a state of chaotic
rotation. An important problem is whether there exists a preferred
orientation of the satellites in chaotic rotation, or their
``chaotic tumbling'' is isotropic? This is important for drawing
conclusions about the character of rotation from observational
data.

For describing rotation of a satellite we use a set of Euler
angles adopted in \citep{W87}. It is different from that used
above. The reason for the change is that the anisotropy of
orientation with respect to the direction to the planet is
described straightforwardly in the new set. The difference between
the old set and the new one consists in the sequence of imaginary
rotations by the angles from the initial position (identical to
that in the old system; see above) to the actual orientation of
the satellite. In the new set, the rotation is made first by
$\theta$ about $c$, second, by $\phi$ about $b$, third, by $-\psi$
about $a$, until the axes of inertia of the satellite coincide
with the actual orientation.

Therefore the angle $\phi$ in the new set is the angle between the
largest axis of satellite's figure (the axis of the minimum moment
of inertia) and the orbit plane, and the angle ($\theta - f$) is
the angle between the direction to the planet and the plane
containing the largest axis of satellite's figure and orthogonal
to the orbit plane.

\begin{figure}
\begin{tabular}{c}
{\bf a)} \includegraphics[width=0.75\textwidth]{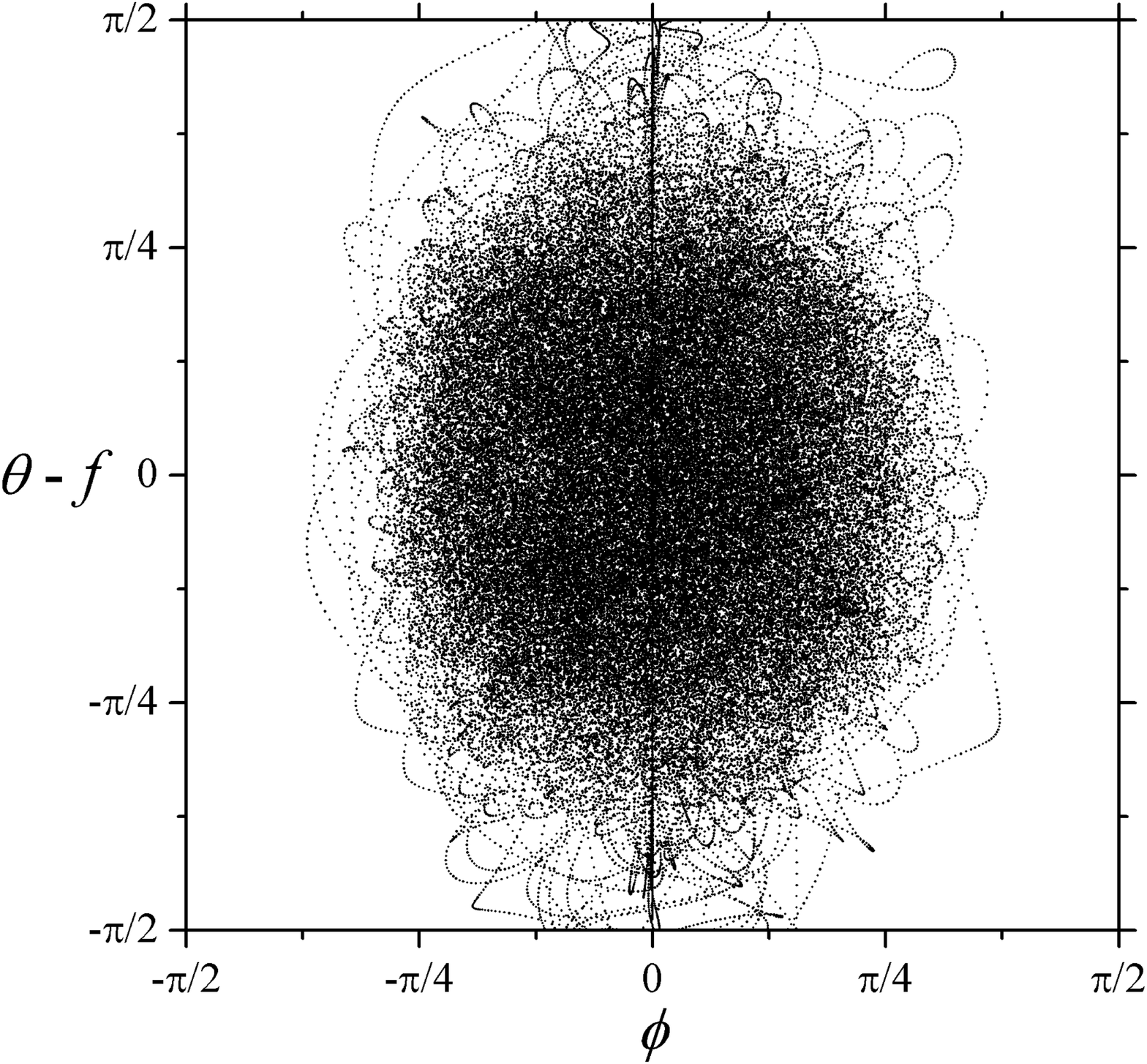}\\
{\bf b)} \includegraphics[width=0.75\textwidth]{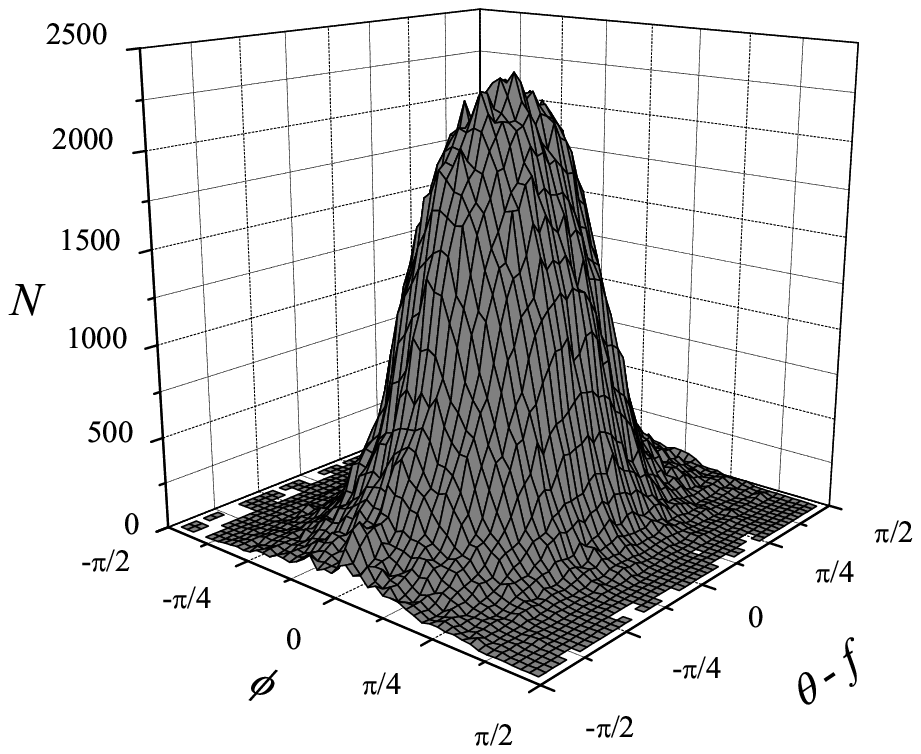}\\
\end{tabular}
\caption{Orientation of Prometheus in chaotic rotation. {\bf (a)}
Projection of the chaotic trajectory to the ($\phi$, $\theta - f$)
plane. The integration time is $1\,000$ orbital periods. {\bf (b)}
A three-dimensional density plot of the discrete projections of
the trajectory to the plane ($\phi$, $\theta - f$). The
integration time is $10\,000$ orbital periods.} \label{fig:9}
\end{figure}

\citet{W87} constructed projection of a chaotic trajectory of
spatial rotational motion of Phobos to the plane ($\phi$, $\theta
- nt$), where $n$ is the orbital mean motion (see Fig.~5 in
\citep{W87}; note that $\theta - f \approx \theta - nt$ for small
eccentricities). The rotation of Phobos with the model initial
conditions close to the separatrices of the 1:2 spin-orbit
resonance was considered. Planar rotation of Phobos is unstable
with respect to tilting the axis of rotation not only near the
separatrices of this resonance, but also practically in the whole
1:2 resonance zone in the phase space \citep{W87}. Fig.~5 in
\citep{W87} shows that spatial rotation of Phobos with such
initial conditions is not totally chaotic: there is a preferred
orientation of the largest axis of satellite's figure in the
direction to the planet. Let us consider an analogous graph for
the chaotic motion of Prometheus close to synchronous 1:1
resonance. In the computation we adopt the values of $b/a$ and
$c/b$ as given in~\citet{EA99} (see Table~1). The resulting
projection of the spatial chaotic trajectory to the plane ($\phi$,
$\theta - f$) is shown in Fig.~\ref{fig:9}(a).

Besides, in Fig.~\ref{fig:9}(b), we build a three-dimensional
density plot of the discrete projections of the trajectory to the
plane ($\phi$, $\theta - f$). The output time step is taken equal
to $0.01$ of the orbital period. The square $(\phi$, $\theta - f)
\in (-\pi/2, \pi/2) \times (-\pi/2, \pi/2)$ is divided in a grid
of $40 \times 40$ pixels. The quantity $N$ designates the number
of the trajectory output points in a given pixel.

The initial data is taken inside the chaotic layer of the
resonance in the phase space section of planar rotation. The graph
in Fig.~\ref{fig:9}(a) is built for a trajectory on the time
interval of $1\,000$ orbital periods, and that in
Fig.~\ref{fig:9}(b) is built for the same trajectory on the time
interval of $10\,000$ orbital periods. We present the graphs only
for Prometheus; in the case of Pandora they look very similar.

The semimajor axes of the orbits of Prometheus and Pandora are
equal to $139\,400$~km and $141\,700$~km, respectively
\citep{F03}; the mean radius of Saturn is equal to $57\,600$~km.
It follows then that the relative area of Saturn's disk as seen
from the satellite (with respect to the area of the celestial
hemisphere) is $8.9$\% for Prometheus and $8.6$\% for Pandora.
These values give the average relative time that the largest axes
of figures of these satellites would be oriented in the direction
to Saturn, if orientations of the satellites during the chaotic
``tumbling'' were isotropic. For the chaotic trajectory presented
in Fig.~\ref{fig:9} we have calculated the values of the average
relative time of orientation towards Saturn; they have turned out
to be equal to $\approx 30$\% for Prometheus and $\approx 22$\%
for Pandora for the time interval of integration of $10\,000$
orbital periods, i.e., the ``isotropic norm'' is exceeded $3.3$
and $2.6$ times, respectively. From the plots in
Figs.~\ref{fig:9}(a) and (b) and these numerical estimates it is
clear that there exists preferred orientation of the largest
satellites' axes in the direction to Saturn, at least for the
given test trajectory.

More extensive additional test computations show that for the used
time interval of integration of $10\,000$ orbital periods the
values of the average relative time of orientation towards Saturn
depend on the choice of initial data. The obtained values are in
the range of 20--30\%. The deviations may indicate that the
observed anisotropy is a temporary effect due to specific initial
conditions, and long term diffusion leads to its disappearance.
This remains an open problem.

\section{Conclusions}
\label{sec:concl}

We have studied possible rotation states of two small moons of
Saturn, Prometheus and Pandora. There are two different regimes of
synchronous rotation in the phase space of planar rotational
motion of Prometheus: $\alpha$-resonance and $\beta$-resonance,
i.e., it is subject to the ``Amalthea effect''. Pandora has
$\alpha$-resonance only. Our analysis of stability of planar
rotation of these satellites with respect to tilting the axis of
rotation has shown that $\alpha$-resonance for Prometheus is
unstable or close to instability, i.e., the satellite most
probably cannot reside in the given regime of synchronous
rotation. Rotation of Prometheus in $\beta$-resonance, as well as
rotation of Pandora in its only possible $\alpha$-resonance, is
close to the attitude instability. Both satellites also possess
period-doubling bifurcation mode of $\alpha$-resonance in the
phase space of rotation. Rotation of both Prometheus and Pandora
in this mode is attitude unstable. With respect to multiplicity of
synchronous states in the phase space, Prometheus is unique
amongst the satellites with known inertial and orbital parameters.
So, whether it is in chaotic rotation or not, its potential
rotational dynamics are rich and complicated. To a less extent the
same is true for Pandora.

Our analysis of the attitude stability of planar rotation of
Prometheus and Pandora for trajectories of various sort (periodic,
quasiperiodic, chaotic) on a representative set of initial data,
carried out by means of computation of the Lyapunov spectra, has
shown presence of alternating concentric ring-like zones of stable
and unstable trajectories around the major synchronous states for
both satellites. Hypothetically, the belts of attitude instability
might form ``barriers'' for capturing the satellites in
synchronous rotation.

In a numerical experiment we have demonstrated that the satellites
in chaotic rotation can mimic ordinary regular synchronous
behaviour: they can have preferred orientation for long periods of
time, the largest axis of satellite's figure being directed
approximately towards the planet. The presence of such anisotropy
of orientation of the satellites in chaotic rotation might prevent
clearing up the character of rotation in observations. Whether
this anisotropy is a temporary effect due to specific initial
conditions, with long term diffusion leading to its disappearance,
remains an open problem.

\end{document}